\documentclass[physrev,reprint,bibnotes]{revtex4-2}  

\usepackage[T1]{fontenc}
\usepackage[australian]{babel}\usepackage[a4paper,centering,hmargin=1.75cm,vmargin=2cm]{geometry} 

\usepackage{amsmath,amssymb,graphicx,bm,microtype} 
\usepackage[dvipsnames]{xcolor} 
\usepackage{booktabs} 
\usepackage[colorlinks,allcolors=blue!50!black]{hyperref} 
\usepackage[all]{hypcap} 
\usepackage{cleveref} 

\usepackage{braket} 
\usepackage[version=4]{mhchem} 
\usepackage{upgreek,textgreek} 
\usepackage{siunitx,textcomp} 
\renewcommand{\d}{\mathop{}\!d} 
\newcommand{\dg}{\dagger}

\newcommand{\hc}{\mathrm{h.c.}}
\newcommand{\D}{\mathrm{D}}
\newcommand{\A}{\mathrm{A}}
\newcommand{\E}{\mathrm{E}}
\newcommand{\pha}[1]{^{\phantom{#1}}}
\begin{document}

\title{Engineering Quantum-Enhanced Transport by Supertransfer}

\author{Adesh Kushwaha}
\affiliation{School of Chemistry, University of Sydney, NSW 2006, Australia}
\author{Ivan Kassal}
\email[]{ivan.kassal@sydney.edu.au}
\affiliation{School of Chemistry, University of Sydney, NSW 2006, Australia}


\begin{abstract}
Collective behaviour of the components of a quantum system can significantly alter the rates of dynamical processes within the system. A paradigmatic collective effect is superradiance, the enhancement in the rate that radiation is emitted by a group of emitters relative to that emitted by independent emitters. Less studied are collective effects in energy transport, notably supertransfer, the enhancement of the rate of energy transfer from donors to acceptors due to delocalised excitations. Despite its generality as a collective enhancement mechanism in transport, there has been no direct experimental detection of supertransfer because, in most physical platforms where delocalisation arises naturally, delocalisation cannot be turned on and off to evaluate its effect on energy transfer. Here, we show that supertransfer could be directly observed using a quantum device based on a superconducting circuit. The programmability and control offered by an engineered device would allow controllable delocalisation of quantum states, giving full tunability over supertransfer. Our guidelines for engineering supertransfer could inform the design of future quantum-enhanced devices.
\end{abstract}

\maketitle

Collective (or cooperative) effects occur when the rates of dynamical processes in a many-body system depend on its components' interactions with each other.
The first collective effect to be studied was Dicke superradiance, where the delocalisation of the system's wavefunction across many atoms increases the rate of emission~\cite{Dicke1954,Gross1982}.
The collective effect that is the opposite of superradiance is superabsorption, where delocalisation over many particles enhances the absorption rate of radiation~\cite{Higgins2014,Quach2022}, a feature that is exploited in proposed quantum batteries~\cite{Quach2020,Quach2022,Quach2023}.

Collective effects also occur in transport, where supertransfer is the general phenomenon of collective enhancement of transport rates due to delocalisation~\cite{Strk1977,Sumi1998,Lloyd2010,Merkli2012,Baghbanzadeh2016,kassal23,Kushwaha2025CEEAT}. Theoretical frameworks that capture this enhancement include generalised and multichromophoric F{\"o}rster resonant energy transfer (gFRET and MC-FRET)~\cite{Mukai1999,Jang2004} in non-radiative energy transfer and generalised Marcus theory (gMT)~\cite{kassal23} in electron transfer. The rate enhancement due to supertransfer provides a general mechanism by which coherent delocalisation can improve transport efficiency~\cite{Tomasi_2020}. Such delocalised excitations arise naturally in molecular aggregates~\cite{Mukai1999,Scholes2000,Scholes2003,Baghbanzadeh2016,Baghbanzadeh2016_PCCP} and in organic photovoltaic (OPV) materials~\cite{Abramavicius2016,Gluchowski2018,Balzer2022,Balzer2023,Balzer2024}, where electronic coupling between neighbouring molecules can produce extended excitonic states.

In most cases of supertransfer, transfer takes place predominantly in the single-excitation regime, where only one excitation is shared among many sites. In this limit, supertransfer arises from delocalisation of a single excitation across an aggregate. Aggregates supporting multiple excitations can exhibit stronger, superradiant-like scaling of the transfer rate~\cite{Lloyd2010}. In this work, we adopt the general definition of collective transport~\cite{Kushwaha2025CEEAT}, where any deviation of the transfer rate from that of independent sites is regarded as collective.

Delocalisation can also suppress transfer, a phenomenon known as subtransfer. Here, the same coherent effects that enhance transport in supertransfer instead reduce the rate below that of independent sites, possibly down to zero. Such suppression can be advantageous in contexts where isolation from loss channels or protection against unwanted energy flow is desired~\cite{Tomasi_2020}.

Despite the promise to enhance transfer efficiency, there have been no direct experimental demonstrations of supertransfer or subtransfer. In systems where delocalisation occurs naturally (e.g., molecular aggregates), this is likely due to the impossibility of turning coherent effects on and off, making it impossible to compare normal transfer and supertransfer in the same system to show a clear enhancement due to delocalisation. 

Here, we propose an artificial quantum device that allows tuning the delocalisation in a way that could be used to definitively demonstrate supertransfer and subtransfer. Our proposal is based on superconducting circuits, which offer programmability and precise control for studying complex quantum phenomena~\cite{Potonik2018,Blais2021,Kim2022}. Furthermore, we present design guidelines for engineering these collective effects, which could inform the development of future quantum-enhanced devices.

\section{Supertransfer}

Supertransfer \cite{Strk1977} is the enhancement of the transfer rate from an aggregate (or group) of donor sites to an aggregate of acceptor sites due to coherent delocalisation among strongly interacting donors, acceptors, or both. In the following, we will use the language of energy transfer, where the donors are initially excited and transfer excitation energy (or excitons) to the acceptors; however, our results hold true for other types of supertransfer, for example of charge.
The transfer rate $\gamma$ is defined by the rate of change of the acceptor population $P_\A$,
    \begin{equation}
        \label{eqn:gamma_tr}
        \frac{dP_\A}{dt}=\gamma P_\D,
    \end{equation}
where $P_\D$ is the population of the donor aggregate.
    
The rate $\gamma$ can be derived through time-dependent perturbation theory or Fermi's golden rule~\cite{May2011} when the coupling between the donor and acceptor aggregates is weak compared to the system-environment coupling. We assume that donor and acceptor sites are coupled through $H_\mathrm{DA}=\sum_{jk}V^{\D\A}_{jk}\ket{A_k}\bra{D_j}+\hc$, where the states $\ket{D_j}$ and $\ket{A_k}$ describe an excitation on an individual donor or acceptor site. The rate of transfer from donor eigenstate $\ket{i}$ to acceptor eigenstate $\ket{f}$ is then
    \begin{equation}
        \label{eqn:Fermi_rule}
        \gamma=\frac{2\pi}{\hbar}\left|\bra{i}H_\mathrm{DA}\ket{f}\right|^2\nu(\omega),
    \end{equation}
where $\nu(\omega)$ is the density of states near $\ket{f}$. 

Normal transfer occurs when both eigenstates $\ket{i}$ and $\ket{f}$ are localised on individual sites, which means that each donor site independently transfers energy to acceptor sites. The transfer rate from localised state $\ket{i}=\ket{D_j}$ to localised state $\ket{f}=\ket{A_k}$ is 
\begin{equation}
    \gamma_\mathrm{loc}\propto \left|\bra{A_k}H_\mathrm{DA}\ket{D_j}\right |^2.
\end{equation}
The same result is obtained if the initial donor state is in the mixed state $\sum_j p_j \ket{D_j}\bra{D_j}$, assuming that all the $V^{\D\A}_{jk}$ are equal. In that case, the rate is
\begin{equation}
    \gamma_\mathrm{mix}\propto \sum_j p_j\left|\bra{A_k}H_\mathrm{DA}\ket{D_j}\right |^2 = \gamma_\mathrm{loc},
\end{equation}
assuming $\nu(\omega)$ remains constant.

By contrast, supertransfer occurs between delocalised eigenstates. To show the enhancement due to delocalisation, we consider an excited delocalised eigenstate of the donor aggregate,
    \begin{equation}
        \label{eqn:Delocalised_Donors}
        \ket{D_\alpha}=\sum_{j=1}^{N_\D}c_{\alpha j}\ket{D_j},
    \end{equation}
where $N_\D$ is the number of donor sites.
For example, in a fully delocalised donor state, $c_{\alpha j}=1/{\sqrt{N_\D}}$, while a localised state ($\ket{D_{j'}}$) has $c_{\alpha j}=\delta_{jj'}$.
Similarly, we take a delocalised eigenstate of the acceptor aggregate,
        \begin{equation}
            \ket{A_\beta}=\sum_{k=1}^{N_\A}c_{\beta k}\ket{A_k}.
        \end{equation}

\begin{figure}[t]
    	\centering
    	\includegraphics[width=\columnwidth]{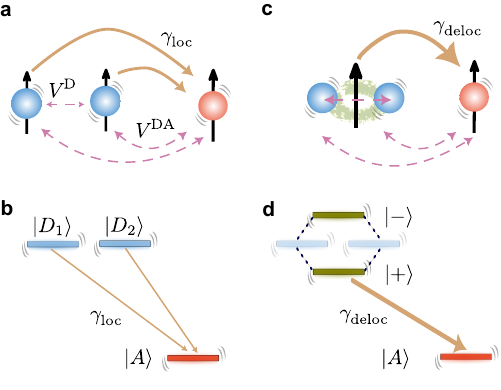}
    	\caption{
    	\textbf{Simplest example of supertransfer.} Two degenerate donor sites (blue) interact through a dipole-dipole interaction (purple) with each other and with an acceptor site (red). \textbf{a) Normal transfer.}~An excitation transfers from weakly coupled donor sites to the acceptor. Because of their weak coupling, the donors are in a mixed state of $\ket{D_1}$ and $\ket{D_2}$, and transfer to $\ket{A}$ occurs at a rate of $\gamma_\mathrm{loc}$. \textbf{b}) The energy diagram corresponding to (a). \textbf{c) Supertransfer.} The excitation transfers from strongly coupled donors in the delocalised eigenstate $\ket{+}=\left (\ket{D_1}+\ket{D_2} \right )/\sqrt{2}$, with a rate $\gamma_\mathrm{deloc}$. The transition dipole of $\ket{+}$ is larger than the transition dipoles of local states, leading to faster transfer, $\gamma_\mathrm{deloc}=2\gamma_\mathrm{loc}$. \textbf{d}) The energy diagram corresponding to (c).
    	}
    	\label{fig:Model}
\end{figure}

The transfer rate between the delocalised eigenstates $\ket{i}=\ket{D_\alpha}$ and $\ket{f}=\ket{A_\beta}$ is 
\begin{equation}
    \gamma_\mathrm{deloc}\propto \left|\bra{A_\beta}H_\mathrm{DA}\ket{D_\alpha}\right |^2=\Bigg|\sum_{j,k}c_{\alpha j}c^*_{\beta k}V^{\D\A}_{jk}\Bigg|^2.
    \label{eqn:interfernce}
    \end{equation}
This rate is maximised for completely delocalised eigenstates ($c_{\alpha j}=1/\sqrt{N_\D}$ and $c_{\beta k}=1/\sqrt{N_\A}$),

\begin{equation}
        \gamma_\mathrm{deloc}^\mathrm{max}=\gamma_\mathrm{loc}\Bigg|\sum_{j}\sum_{k}\frac{1}{\sqrt{N_\D}}\frac{1}{\sqrt{N_\A}}\Bigg|^2=N_\D N_\A\gamma_\mathrm{loc},
        \label{eqn:gamma_deloc}
    \end{equation}
meaning that delocalisation can enhance the transfer rate by a factor of $N_\D N_\A$ over the normal transfer rate~\cite{Strk1977,Lloyd2010}.

Subtransfer---the suppression of transport due to delocalisation---arises if the coefficients $c_{\alpha j}$ and $c_{\beta k}$ in~\cref{eqn:interfernce} carry relative phases that cause destructive interference and reduce the transfer rate below $\gamma_\mathrm{loc}$, including, for complete destructive interference, to zero. An eigenstate corresponding to this cancellation is referred to as a dark state, whereas an eigenstate that produces enhanced transfer is a bright state. 

In general, we describe as supertransfer any increase in the transfer rate~\cite{Kushwaha2025CEEAT}, i.e., any situation where
    \begin{equation}
        \gamma_{\mathrm{loc}}<\gamma_{\mathrm{deloc}}\le N_\D N_\A\gamma_{\mathrm{loc}},
        \label{eqn:enhanc_gen}
    \end{equation} 
and as subtransfer any decrease, $0\le\gamma_\mathrm{deloc}<\gamma_\mathrm{loc}$. The minimal system exhibiting supertransfer requires $N_\D=2$ and $N_\A=1$ (\cref{fig:Model}), resulting in $\gamma_{\mathrm{loc}}\le \gamma_{\mathrm{deloc}}\le 2\gamma_{\mathrm{loc}}$.

The transfer rate can increase even further if the number of excitations exceeds the previously assumed value of one. In that case, the maximum rate scales with the size of the system as $O(N_\D^2N^{\phantom{2}}_\A)$~\cite{Lloyd2010}. For $N_\mathrm{exc}$ donor excitations, the supertransfer rate~\cite{Lloyd2010,Kushwaha2025CEEAT} is
  \begin{equation}
        \gamma^\mathrm{max}_\mathrm{deloc}=N_\mathrm{exc}(N_\D-N_\mathrm{exc}+1)N_\A\gamma_0.
        \label{eqn:ST_rate}
    \end{equation} 
Here, we initially focus on the single-excitation case, because it provides the simplest demonstration of supertransfer, before generalising our results to the multi-excitation case.

\section{Model}
To construct a demonstration of supertransfer, we consider aggregates of $N_\mathrm{D}$ donors and $N_\mathrm{A}$ acceptors, which are all coupled to each other and to an environment, with the Hamiltonian
\begin{equation}
    \label{eqn:H_full}
    H=H_\D+H_\A+H_{\D\A}+H_{\D\E}+H_{\A\E}+H_{\E},
\end{equation}
containing terms describing the donors (D), the acceptors (A), the environment (E) and their interactions (mixed indices). To model the simplest case of supertransfer, we assume only a single excitation in the system and model each donor and acceptor site as a two-level system. Donor sites $D_j$ have energies $E^\D_{j}$ and interact through couplings $V^\D_{jj'}$, 
\begin{equation}
    \label{eqn:H_donor}   H_\D=\sum_{j}E^\D_{j}\ket{D_{j}}\bra{D_{j}}+\sum_{j\neq j'}V^\D_{jj'}\ket{D_{j}}\bra{D_{j'}}.
\end{equation}
Similarly, in the acceptor site basis,
\begin{equation}
    \label{eqn:H_acceptor}   H_\A=\sum_{k}E^\A_{k}\ket{A_{k}}\bra{A_{k}}+\sum_{k\neq k'}V^\A_{kk'}\ket{A_{k}}\bra{A_{k'}}.
\end{equation}
Donor and acceptor sites interact with each other through
\begin{equation}
    \label{eqn:H_int} 
    H_\mathrm{DA}= \sum_{j,k}V^{\D\A}_{jk} \ket{D_{j}}\bra{A_{k}}+\hc
\end{equation}
The environment is modelled by coupling each site to an independent bath of harmonic oscillators,
\begin{equation}
    \label{eqn:H_bath}
    H_\E=\sum_{j}\sum_{\xi}\omega^{\D}_{j\xi}a_{j\xi}^{\D\dagger} a^{\D}_{j\xi}+\sum_{k}\sum_{\xi}\omega^{\A}_{k\xi}a_{k\xi}^{\A\dagger} a^{\A}_{k\xi},
\end{equation}
where $\hbar=1$ and $a^{\D}_{j\xi}$ is the annihilation operator for the $\xi$th environment mode of donor $j$. Similarly, $a^{A}_{k\xi}$ is the operator for the $\xi$th mode on acceptor $k$. The system-environment interactions are assumed to be linear,
\begin{align}
    \label{H_bath_donor}
    H_{\D\E}&=\sum_{j}\sum_{\xi}g^{\D}_{j\xi}(a^{\D}_{j\xi}+a_{j\xi}^{\D\dagger})\ket{D\pha{\dg}_{j}}\bra{D\pha{\dg}_{j}},\\
    \label{H_bath_acceptor}
    H_{\A\E}&=\sum_{k}\sum_{\xi}g^{\A}_{k\xi}(a^{\A}_{k\xi}+a_{k\xi}^{\A\dagger})\ket{A\pha{\dg}_{k}}\bra{A\pha{\dg}_{k}}.
\end{align}
The system-environment interactions can be specified by their spectral densities, $J^{\D}_j(\omega)$ for donors and $J_k^{\A}(\omega)$ for acceptors, where
\begin{equation}
    \label{eqn:J_omega}
    J^{\D}_j(\omega)=\sum_{\xi}{(g^{\D}_{j\xi}})^2\delta(\omega-\omega_{\xi})
\end{equation}
and similarly for $J_k^{\A}(\omega)$. The strength of system-environment coupling is quantified by the reorganisation energy $\lambda$, defined for the $j$th donor as
\begin{equation}
    \label{eqn:Reorgan}
    \lambda_j^\D=\int_0^\infty \frac{J^{\D}_j(\omega)}{\omega}\d\omega,
\end{equation}
and similarly for acceptors. For simplicity, we assume the reorganisation energies of the donors are the same ($\lambda_j^\D=\lambda^\D$) and similarly for the acceptors ($\lambda_k^\A=\lambda^\A$).

The simplest model of supertransfer can be obtained using classical noise, which manifests itself as time-dependent fluctuations of site energies: $E^\D_{j} \to E^\D_{j}+\delta E^\D_{j}(t)$, $E^\A_{k} \to E^\A_{k}+\delta E^\A_{k}(t)$.
Classical noise in site energies induces dephasing in the site basis, whose effect on transfer~\cite{Potonik2018,Huang2020} depends on its spectral density $J(\omega)$; here, we use, at each site,  the Drude-Lorentz spectral density
\begin{equation}
    \label{eqn:Lorentz_Drude}
    J(\omega)=\frac{2\omega\lambda}{\omega^2+\omega_c^2},
\end{equation}
where $\omega_c$ is a cutoff frequency. Additionally, we assume an $\omega_c$ large compared to system frequencies, which yields Markovian dephasing.

The initial eigenstate is the lowest-energy state of the donor aggregate, whose extent of delocalisation varies based on system parameters. By tuning these parameters, the donor aggregate can be prepared in either a fully localised or a fully delocalised state (\cref{fig:Model}). 
To enact normal transfer, excitation is localised on a donor site, or, more generally, in a statistical mixture of donor sites in the density matrix
$\sum_{j=1}p_j\ket{D_j}\bra{D_j}$.
By contrast, to enact supertransfer, the excitation starts in a delocalised state $\ket{D_\alpha}$.

\begin{figure}[t]
    \includegraphics{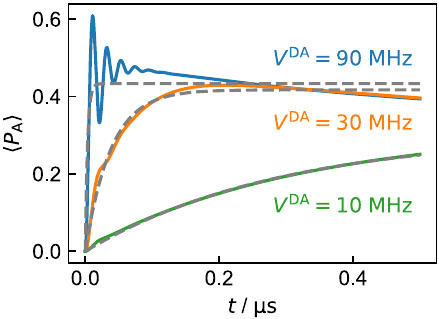}
    \caption{
    \textbf{Evaluating the transfer rate $\gamma$ from population dynamics.} Population transfer to acceptor plotted for three values of donor-acceptor coupling $V^\mathrm{DA}$ with $\lambda^\A=\qty{80}{\mega Hz}$ and $\lambda^\D=\qty{10}{\mega Hz}$. The solid lines are the numerically exact calculations of the acceptor populations, which are fitted to~\cref{eqn:gamma_tr_const} (dashed lines) to find the values of $\gamma$. The fit to the acceptor population is meaningful only in the weak-coupling regime where Rule~1 applies. The transfer is described by a rate equation when Rule~1 ($V^{\D\A} \ll \lambda^\D+\lambda^\A$) is met, as it is for $V^\mathrm{DA}=\qty{10}{\mega Hz}$. At $V^\mathrm{DA}=\qty{30}{\mega Hz}$, the exponential behaviour begins to break down, and it is a completely inaccurate description of the dynamics at $V^{\D\A}=\qty{90}{\mega Hz}$. Other parameter values are given in~\cref{tab:parameters}. Calculations performed using QuTip~\cite{Johansson_2013}.} 
    \label{fig:Pop_acceptor}
\end{figure}

We determine $\gamma$ by numerical simulation of the dynamics, generated by $H$, of the system's reduced density matrix $\rho(t)$ (\cref{fig:Pop_acceptor}). In the limit of weak donor-acceptor coupling, the dynamics is described by the rate equation in~\cref{eqn:gamma_tr} and $\gamma$ is given by the golden-rule expression in \cref{eqn:Fermi_rule}. To extract $\gamma$ from the dynamics, we fit the acceptor population $P_\A(t)$ to
\begin{equation}
    \label{eqn:gamma_tr_const}
    P_\A(t)=P_\A^{\infty}(1-e^{-\gamma t}),
\end{equation}
where $P_\A(t)=\sum_{j}\bra{A_j}\rho(t)\ket{A_j}$ and $P_\A^{\infty}$ is the final acceptor population.

\begin{figure*}[t]
    \centering
    \includegraphics[width=\textwidth]{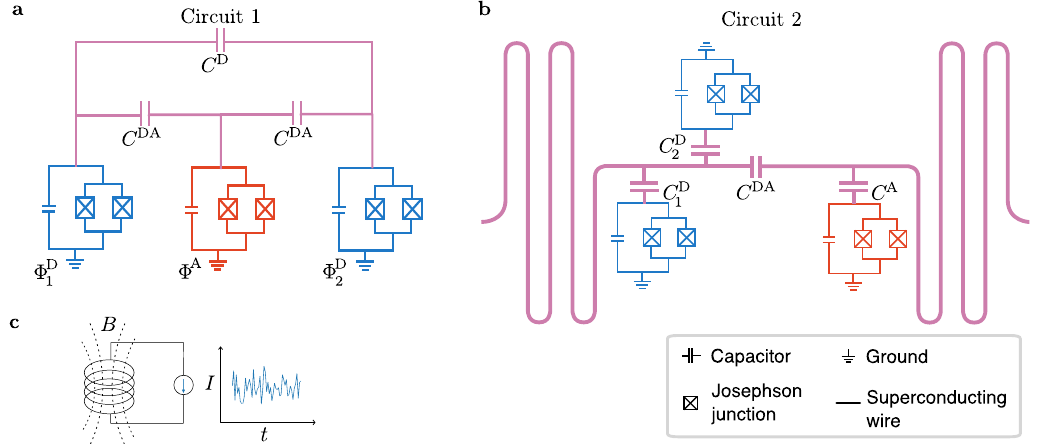}
 \caption{
    \textbf{Proposed superconducting circuit quantum electrodynamics (cQED) devices demonstrating controllable supertransfer.} Both circuits realise the model of \cref{fig:Model} using flux-tunable transmon qubits. \textbf{a)}~In Circuit~1, the donors (blue) are coupled to each other directly through capacitors. \textbf{b)}~In Circuit~2, the coupling between sites is mediated by two buses (purple). \textbf{c)} In both circuits, noise can be introduced using an independent solenoid on each site; fluctuating the current $I$ through the solenoids produces fluctuating magnetic fluxes $\Phi$ through the transmon loops, changing their site energies. }
    \label{fig:Quantum_Simulator}
\end{figure*}

\section{Design Rules for Supertransfer and Subtransfer}

There are two design rules for collective effects in transport (supertransfer and subtransfer), both of which take the form of time-scale separations. 

\textit{Rule 1:} the transfer must be well described by the golden-rule rate process of \cref{eqn:Fermi_rule}. This requirement is met if the donor aggregate is weakly coupled to the acceptor aggregate, compared to the system-environment coupling quantified by the reorganisation energies,
\begin{equation}
    \label{eqn:Rule1}
    |V^{\D\A}| \ll \lambda^\D+\lambda^\A.
\end{equation} 

\textit{Rule 2:} the eigenstates of either the donor or the acceptor aggregates (or both) must be delocalised. For the donors (with the same applying for the acceptors), this can be achieved if the intra-aggregate couplings $V^{\D}$ are greater than both the static disorder in the aggregate (the variance $\delta^\D$ of the set $\{E^\D_j\}$) and the dynamic disorder (quantified by $\lambda^\D$), i.e.,
\begin{equation}
\label{eqn:Rule2}
    V^{\D} \gtrsim \delta^\D, \lambda^\D.
\end{equation}

The maximum possible supertransfer, with rate $\gamma_{\mathrm{deloc}} = N_\D N_\A\gamma_{\mathrm{loc}}$, is achieved if the inequality in \cref{eqn:Rule2} is taken to the limit $V^{\D} \gg \delta^\D, \lambda^\D$. Similarly, complete inhibition of transfer is achieved in the limit  $V^{\D} \gg \delta^\D, \lambda^\D$.

\section{Experimental proposal}

\begin{table*}[t]
    \centering
    \begin{tabular}{lllllS[table-format=4.2]}
        \toprule
         Parameter           & Value (\qty{}{\mega\hertz})              & Implementation in        & Implementation in \\
         \phantom{Parameters} & \phantom{Value (\qty{}{\mega\hertz})}    & Circuit 1                & Circuit 2          \\
        \midrule
         $V^{\D}$            & 10    & $C^{\D}$                                               & $\tilde C^{\D}$                                          & \\
         $V^{\D\A}$          & 10    & $C^{\D\A}$                                             & $\tilde C^{\D\A}$                                       & \\
         $E^\D-E^\A$       & 148   & $E^\D-E^\A$           & $\tilde E^\D-\tilde E^\A$ & \\
       $\delta^{\D}$       & 5     & $\langle \delta E^\mathrm{D}(t)\rangle$                & $\langle \delta \tilde {E}^\mathrm{D}(t)\rangle$                 & \\
         $\lambda^\D$        & 10    & $\lambda^\mathrm{cl}_\D$                 & $\tilde\lambda^\mathrm{cl}_\D$                 & \\         
         $\lambda^\A$        & 80    & $\lambda^\mathrm{cl}_\A$                 & $\tilde\lambda^\mathrm{cl}_\A$                 & \\
        \bottomrule
    \end{tabular}
    \caption{
        \textbf{Sample of proposed parameters} for showing supertransfer in Circuit 1 and Circuit 2, being values that are experimentally feasible in cQED experiments \cite{Blais2021}.
    }
    \label{tab:parameters}
\end{table*}

    The design rules above can be used to demonstrate supertransfer using a superconducting architecture.

    The simplest case of supertransfer---shown in~\cref{fig:Model}---could be implemented by meeting three requirements. First, it requires three sites: two donors and one acceptor ($N_D=2$ and $N_A=1$). Second, the energies of these sites must be tunable to ensure that the acceptor energy is lower than those of the donors and to allow noise injection. Third, it must be possible to control delocalisation to show a higher transfer rate in the delocalised case compared to the localised one.
    
    These requirements are all achievable experimentally using a superconducting circuit quantum electrodynamics (cQED) architecture. In cQED, the sites can be implemented as transmon qubits whose energy can be tuned using an external magnetic field~\cite{Koch2007}. The couplings between the sites can be mediated either by capacitive couplings as in \cref{fig:Quantum_Simulator}a or using a bus as in \cref{fig:Quantum_Simulator}b. 
    
    The Hamiltonian of the required cQED circuit is
         \begin{multline}
            \label{eqn:H_Circuit}
            H_\text{circuit}=E^\D\sigma^{z}_{1}+(E^\D+\delta^{\D})\sigma^{z}_{2}+E^\A\sigma^{z}_{\A}\\
            +\big(C^\mathrm{D}\sigma^{x}_{1}\sigma^{x}_{2}+C^\mathrm{DA}(\sigma^{x}_{1}\sigma^{x}_{\A}+\sigma^{x}_{2}\sigma^{x}_{\A})
            +\hc\big),
        \end{multline}
     where $\sigma^z_i$  and $\sigma^x_i$ are Pauli operators for site $i$, while $C^\mathrm{D}$ and $C^\mathrm{DA}$ are couplings. For example, for donor~1, $\sigma^z_1=\ket{D_1}\bra{D_1}-\ket{G}\bra{G}$ and $\sigma^{x}_{1}=\ket{D_1}\bra{G}+\ket{G}\bra{D_1}$, where $\ket{G}$ is the state of the system with no excitations. $H_\text{circuit}$ preserves the number of excitations and reduces to $H$ when restricted to its single-excitation subspace.

     Site energies (and therefore the static disorder) are tunable because transmon energies are sensitive to external magnetic fields~\cite{DiCarlo2009,Krantz2019,Blais2021},
        \begin{equation}
            \label{eqn:Tunable_qubit}
                H_\mathrm{qb}=E_\mathrm{qb}(\Phi)\sigma_z.
        \end{equation}
        The energy $E_\mathrm{qb}$ can be tuned between hardware-dependent values of $E_\mathrm{min}$ and $E_\mathrm{max}$ by tuning the magnetic flux $\Phi$ through the transmon's superconducting loop~(\cref{fig:Quantum_Simulator}) from 0 to $\Phi_0\pi/2$, where $\Phi_0=h/2e$ is the  flux quantum \cite{Krantz2019}. Experimentally, $E_\mathrm{qb}(\Phi)$ is typically tunable across a range $E_\mathrm{max}-E_\mathrm{min}$ of around $\qty{100}{\mega Hz}$, or, in some hardware, up to $\qty{1}{\giga Hz}$~\cite{Schreier2008,DiCarlo2009,Rol2020}.

        The tunability of flux-tunable transmons is also used to engineer a dephasing environment by fluctuating the energy levels~\cite{Potonik2018}. To generate the fluctuating flux $\Phi(t)$, a fluctuating current $I(t)$ is injected into a solenoid to generate a fluctuating magnetic field $B(t)$ as in~\cref{fig:Quantum_Simulator}c. For small currents, the change in the energy of site $i$ ($\delta E_i(t)=E_i(t)-\langle E_i(t) \rangle$) due to a change in current $\delta I(t)$ is
     \begin{equation}
         \label{eqn:Fluctuation}
         \delta E_i(t) \approx \frac{d E_i}{d \Phi_i}\frac{d \Phi_i}{d B_i}\frac{d B_i}{d I_i}  \delta I_i(t),
     \end{equation}
     where the product of the derivatives is a calibration factor that can be experimentally determined.

     The effect of the dynamic disorder induced by the classical noise signal $\delta E_i(t)$ is determined by its spectrum $J^{\mathrm{cl}}(\omega)$, which is the Fourier transform of the site-energy correlation function,
     \begin{equation}
         \label{eqn:Noise_spectra}
         J^\mathrm{cl}(\omega)=\mathcal{F}\{\langle \delta E_i(t)\delta E_i(0)\rangle\}.
     \end{equation}
     The overall effect of the noise is quantified by the reorganisation energy
        \begin{equation}
            \label{eqn:Noise_lambda}
            \lambda^\mathrm{cl}=\int_0^\infty \frac{J^\mathrm{cl}(\omega)}{\omega}\d\omega.
        \end{equation}
        
\begin{figure}
    \centering
    \includegraphics{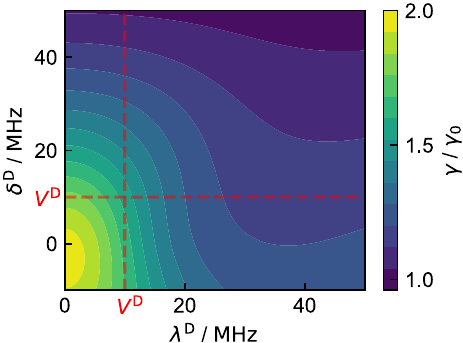}
    \caption{\textbf{Donor parameters in Circuit 1 can be tuned to switch the dynamics from normal transfer at rate $\gamma_0$ to maximum possible supertransfer at rate $2\gamma_0$.} Supertransfer is achieved when Rule~2 is met, i.e., the reorganisation energy $\lambda^\D$ and the site detuning $\delta^\D$ are both smaller than the intra-aggregate coupling $V^{\D}$, as is the case in the bottom-left of the plot. The rate gradually transitions to that of normal transfer as either parameter moves away from this region. The other parameter values are as in \cref{tab:parameters}, giving $\gamma_0=\qty{0.77}{\micro s^{-1}}$.
    }
 \label{fig:donor_donor_parameters}
\end{figure}

    The final component, the controllable couplings between sites, can be realised in two different ways, as shown in \cref{fig:Quantum_Simulator}.
     
    In \textit{Circuit 1} (\cref{fig:Quantum_Simulator}a), the couplings $C$ are implemented as direct capacitive couplings. These can be either fixed \cite{Pashkin2003,Berkley2003,Bialczak2010} or adjustable using a longitudinal electric field~\cite{Wu2018}. Capacitive coupling allows a direct implementation of the desired cQED Hamiltonian of \cref{eqn:H_Circuit}.

    \begin{figure*}
    	\centering
    	\includegraphics[width=\textwidth]{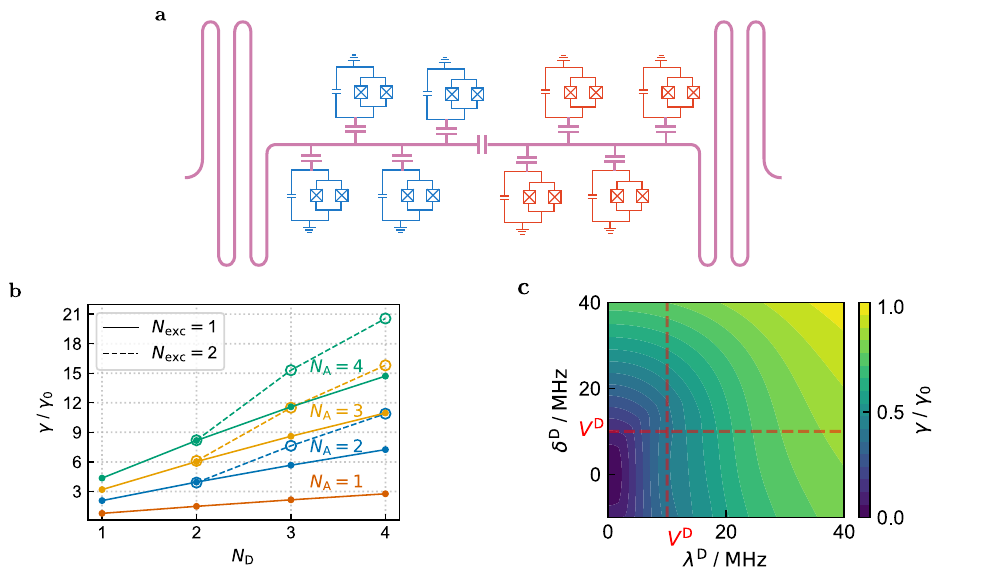}
    	\caption{
    	\textbf{Extending the supertransfer framework to multi-excitation dynamics and subtransfer}. \textbf{a)} Expansion of Circuit~2 for demonstrating the scaling of the transfer rate. The donors are connected through a bus and the acceptors are connected through a second bus capacitively coupled to the first. The number of donors or acceptors participating in the dynamics could be changed on the fly by detuning the energy of unwanted qubits to be much higher than the others. 
        \textbf{b)}~Scaling of supertransfer with the number of donors $N_\mathrm{D}$ and the number of excitations $N_\mathrm{exc}$.
        For a single excitation ($N_\mathrm{exc}=1$), rates are obtained using~\cref{eqn:gamma_tr_const} and scale as $N_\mathrm{D}N_\mathrm{A}$, as expected in \cref{eqn:enhanc_gen}. For two excitations ($N_\mathrm{exc}=2$), the maximum rate is the initial transfer rate of~\cref{eqn:gamma_tr}, and is enhanced further above the single-excitation case, scaling as $2(N_\mathrm{D}-1)N_\mathrm{A}$, as expected in \cref{eqn:ST_rate}. 
        In both cases, for large $N_\mathrm{D}$ and $N_\mathrm{A}$, the rates are slightly smaller than the predicted values because eigenstate energies change slightly as additional sites are added.
        \textbf{c)} Subtransfer, engineered in the $N_\D=2$ and $N_\A=1$ case by tuning the donor parameters $\delta^\D$ and $V^{\D}$. The donors are prepared in the dark eigenstate $\ket{-}$. As in supertransfer, subtransfer is achieved when Rule~2 is met (bottom-left of the plot). The rate increases to that of normal transfer as either parameter moves away from this region. The other parameter values are as in \cref{tab:parameters}, giving $\gamma_0=\qty{0.77}{\micro s^{-1}}$.}
    	\label{fig:Scaling_Circuit}
    \end{figure*}

     By contrast, in \textit{Circuit 2} (\cref{fig:Quantum_Simulator}b), the couplings are mediated by buses~\cite{Blais2004,Majer2007,Tao_2020}. This approach has the advantage of being more easily scalable to greater numbers of donors and acceptors. Bus-mediated coupling can be understood using the Hamiltonian of Circuit~2, where both of the donors are coupled to the donor bus (a microwave cavity resonator), the acceptor is coupled to the acceptor bus, and the two buses are capacitively coupled to each other, 
        \begin{multline}
        \label{eqn:H_circuit2}
            H_2=\sum_{i=1}^2 \left (E^\D_i\sigma^{z}_{i}+C^\D_{i}(b _\D^\dg\sigma^-_{i}+b _\D\sigma_{i}^+)\right) + \omega_{\D} b_\D^\dg b_\D\\
            +E^\A\sigma^{z}_{\A}+C^\mathrm{A}(b_\A^\dg\sigma^-_{\A}+b_\A\sigma_\A^+) + \omega_\A b_\A^\dg b_\A \\
            + C^\mathrm{DA}(b_\D^\dg+b_\D)(b_\A^\dg+b_\A),
        \end{multline} 
    where $\sigma^\pm=(\sigma^x\pm i\sigma^y)/2$, $b_\D$ and $b_\A$ are the donor and acceptor bus annihilation operators, $\omega_\D$ and $\omega_\A$ are the bus resonant frequencies, and $C^\mathrm{D}_i$ and $C^\mathrm{A}$ are the couplings of the qubits to their bus and $C^{\D\A}$ couples the two buses.
    When the sites are far detuned from $\omega_\D$ and $\omega_\A$, the buses can be adiabatically eliminated using the Fr{\"o}hlich-Nakajima transformation~\cite{Blais2007,Tao_2020} to give the effective Hamiltonian
        \begin{multline}
        \label{eqn:H_circuit2eff}
            H_\text{eff}=\sum_{i=1}^{2}\tilde{E}^\D_i\ket{D_i}\bra{D_i} +\tilde{E}^\A\ket{A}\bra{A}+\big(\tilde C^{\D\D}\sigma^-_{1}\sigma_{2}^+\\
            +\tilde C^{1\A}\sigma^-_{1}\sigma_{\A}^++\tilde C^{2\A}\sigma^-_{2}\sigma_{\A}^++\hc\big),
        \end{multline}
    where the effective couplings and energies are
    \begin{align}
    \label{eqn:effective_parameters}
            \tilde C^{\D\D}&=\frac{C^{\D}_{1} }{2 \Delta_{1} }\frac{ C^{\D}_{2}}{2  \Delta_{2}}(\Delta_{1}+\Delta_{2}),\\
            \tilde C^{i\A}&=C^{\D\A}\frac{ C^\D_{i} }{\Delta_{i} }\frac{ C^\A}{\Delta_\A},\\
            \tilde E^\D_i&=2E^\D_i+\frac{(C^\D_{i})^2}{\Delta_{i}},\\
            \tilde E^\A&=2E^\A+\frac{(C^{\A})^2}{\Delta_{\A}},
        \end{align}
    with $\Delta_{i}=2E^\D_{i}-\omega_{\D}$ and $\Delta_{\A}=2E^{\A}-\omega_{\A}$.

    The noise spectrum $\tilde J^\mathrm{cl}(\omega)$ due to fluctuations of $E^\D_i$ and $E^\A$ is given by the Fourier transform of $\langle \delta\tilde E(t)\delta \tilde E(0)\rangle$, and the reorganisation energy $\tilde\lambda^\mathrm{cl}$ by \cref{eqn:Noise_lambda}.
    
    \textit{Simulation results.} 
    We simulate the dynamics in Circuit~1 with the parameters in \cref{tab:parameters} using QuTip~\cite{Johansson_2013}. These parameters are chosen because they meet Rule~1 and Rule~2 for supertransfer and are representative of typical values found in cQED systems \cite{Blais2021}.

    When Rule~1 ($V^{\D\A}\ll\lambda^\D+\lambda^\A$) is met, the transfer is exponential (\cref{fig:Pop_acceptor}). The exponential behaviour breaks down and becomes oscillatory for $V^{\D\A}\gtrsim\lambda$.
    
    When Rule~2 ($V^{\D}\gtrsim\delta^\D,\lambda^\D$) is met, the supertransfer enhances the transfer rate $\gamma$ (\cref{fig:donor_donor_parameters}). The maximum supertransfer rate $2\gamma_0$ (for $N_\D=2$ and $N_\A=1$) gradually decreases to $\gamma_0$ if the donor parameters leave the region defined by Rule~2. This gradual change with the parameters makes supertransfer robust to small changes in either the static ($\delta$) or dynamic ($\lambda$) disorders.

    \textit{Scaling of supertransfer with the number of donors, acceptors, and excitations.}   
    We test the predicted scaling of supertransfer in two stages. First, we consider the single-excitation regime and verify the scaling in \cref{eqn:enhanc_gen}. The numerical results shown in \cref{fig:Scaling_Circuit}b agree with this prediction over the range of donor and acceptor numbers considered.
    Second, we consider multiple excitations and test the more general prediction in \cref{eqn:ST_rate}.
    The results for $N_\mathrm{exc}=2$, also shown in \cref{fig:Scaling_Circuit}b, reproduce the predicted scaling $\gamma^\mathrm{max}_\mathrm{deloc}=2(N_\mathrm{D}-1)N_\mathrm{A}\gamma_0$.
    
    The multi-excitation transfer rate is not constant in time. As excitations move from the donors to the acceptor, the number of excitations remaining on the donors decreases, reducing the collective enhancement available for subsequent transfer. The instantaneous rate therefore falls from its initial maximum value and, once only a single excitation remains on the donor aggregate, approaches the single-excitation rate $N_\mathrm{D}N_\mathrm{A}\gamma_0$. Because the acceptor population is not described by the single exponential as in \cref{eqn:gamma_tr_const}, we determine $\gamma$ from the initial rate defined in \cref{eqn:gamma_tr}, rather than from an exponential fit.
    
    These scaling relations could be tested experimentally using the enlarged version of Circuit~2 shown in \cref{fig:Scaling_Circuit}a. The number of participating donors and acceptors could be varied within a single device by detuning unwanted qubits from the dynamics.

    \textit{Demonstrating subtransfer.}
    The same experimental architecture can also be used to demonstrate subtransfer. By preparing dark eigenstates of the donor or acceptor aggregates, rather than the bright eigenstates used to observe supertransfer, the transfer rate is suppressed below that of independent sites because of destructive interference between the transfer pathways.
    
    The dependence of the subtransfer rate on the reorganisation energy $\lambda^\D$ and static disorder $\delta^\D$ is shown in \cref{fig:Scaling_Circuit}c. As for supertransfer (in \cref{fig:donor_donor_parameters}), the condition for observing the collective effect is Rule~2: the intra-aggregate coupling must dominate both static and dynamic disorder ($V^\D\gtrsim\delta^\D,\lambda^\D$). When this condition is satisfied, the donor eigenstates preserve the phase relationships required for destructive interference, giving the strongest suppression of transfer. Increasing either $\delta^\D$ or $\lambda^\D$ progressively destroys this interference, causing the transfer rate to increase towards that of normal transfer. Like supertransfer, subtransfer is robust to moderate experimental imperfections and can be realised for any sufficiently large intra-aggregate coupling satisfying Rule~2.
    
    In Circuit~2, the required dark eigenstates could be engineered by tuning the effective donor-donor couplings using the bus-mediated architecture. By controlling both the magnitudes and signs of the effective couplings via the qubit-bus detunings, the relative phases within the donor ground state could be adjusted so it becomes dark. 

\section{Discussion}

    We showed that a quantum device based on a cQED circuit could be engineered to conclusively demonstrate supertransfer. The simplest supertransfer case is achieved with $N_\D=2$ and $N_\A=1$, with larger arrangements also possible, as shown in \cref{fig:Scaling_Circuit}. 
    
    Circuits 1 and 2 are achievable with current engineering capabilities in cQED, as all required components have been experimentally demonstrated. The necessary couplings can be implemented using either of two established experimental techniques: direct capacitive coupling~\cite{Pashkin2003,Berkley2003,Bialczak2010} or buses for indirect qubit coupling~\cite{Majer2007,DiCarlo2009,Song2017}. Furthermore, the ability to inject noise into the system has also been demonstrated for both the Drude-Lorentz spectrum we use and for more general ones~\cite{Solinas2012,Potonik2018,Sung2021}. Finally, our proposal can be implemented using simulation times that are an order of magnitude shorter than typical transmon decoherence times of 50--$\qty{120}{\micro s}$~\cite{Blais2021}. 

    The two proposed implementations---direct coupling in Circuit 1 and indirect coupling in Circuit 2---have complementary advantages. In Circuit 1, there is complete flexibility in selecting all of the couplings, which means that the mapping of the supertransfer Hamiltonian is direct and straightforward. However, the implementation of all-to-all connectivity would be challenging when Circuit~1 is scaled up to more donors and acceptors. In contrast, in Circuit~2, the couplings are achieved through a bus, facilitating scalability as illustrated in~\cref{fig:Scaling_Circuit}a. However, the scalability comes at the cost of a less straightforward mapping of the supertransfer Hamiltonian and limited flexibility in tuning the parameters, because tuning one circuit parameter can affect multiple parameters in the effective Hamiltonian (see \cref{eqn:H_circuit2eff}). Circuit~2 has the further advantage that its detuning-based couplings allow for negative couplings, which opens possibilities such as rearranging the order of bright and dark states and facilitating a controlled demonstration of subtransfer.
    
    The robustness of supertransfer and subtransfer to experimental noise simplifies implementation by reducing the need for precise parameter engineering. It is feasible in any system provided that conditions specified by Rule~1 and Rule~2 are satisfied. Furthermore, the resilience to variations in system parameters and environmental conditions may enable applications of supertransfer in quantum-enhanced engineered systems.

\section*{Acknowledgements} 
    We thank Francesco Campaioli, Jared Cole, Xanthe Croot, and Alexander Grimm for valuable discussions. We were supported by the Australian Research Council (DP220103584 and FT230100653) and a Sydney Quantum Academy scholarship.

\bibliography{bib}

\end{document}